\documentclass[letter]{aa} % for the letters 
\usepackage{graphicx}
\usepackage{txfonts}
\usepackage{pdflscape}
\usepackage[colorlinks=true]{hyperref}
\hypersetup{linkcolor=blue, citecolor=blue}
\usepackage{gensymb}

\usepackage{amstext}

\begin{document} 

   \title{Local Bubble contribution to the $353$-GHz dust polarized emission}

   \author{R. Skalidis
          \inst{1} \fnmsep \inst{2} \fnmsep \thanks{rskalidis@physics.uoc.gr}
          \and
          V. Pelgrims \inst{1} \fnmsep \inst{2} \fnmsep 
          \thanks{pelgrims@physics.uoc.gr}
           }

   \institute{Department of Physics, Univ. of Crete, Heraklion 70013, Greece
              \and
             Institute of Astrophysics, FORTH, Heraklion 71110, Greece}

    \date{Received August 21, 2019; accepted October 21, 2019}

  \abstract
{It has not been shown so far whether the diffuse Galactic polarized emission at frequencies relevant for cosmic microwave background (CMB) studies originates from nearby or more distant regions of our Galaxy. This questions previous attempts that have been made to constrain magnetic field models at local and large scales. The scope of this work is to investigate and quantify the contribution of the dusty and magnetized local interstellar medium to the observed emission that is polarized by thermal dust. We used stars as distance candles and probed the line-of-sight submillimeter polarization properties by comparing the emission that is polarized by thermal dust at submillimeter wavelengths and the optical polarization caused by starlight. We provide statistically robust evidence that at high Galactic latitudes ($|b| \geq 60^\circ$), the $353$ GHz polarized sky as observed by \textit{Planck} is dominated by a close-by magnetized structure that extends between $200$ and $300$ pc and coincides with the shell of the Local Bubble. Our result will assist modeling the magnetic field of the Local Bubble and characterizing the CMB Galactic foregrounds.}


\maketitle%
%-------------------------------------------------------------------

\section{Introduction}

Several phenomena can be used to model and constrain the magnetic field of our Galaxy. Dedicated surveys of the cosmic microwave background (CMB) have provided large-coverage and high-quality data sets for studying and better understanding the physics of the magnetized interstellar medium (ISM). Synchrotron and thermal dust diffuse emission has been used to constrain parametric models of the large-scale Galactic magnetic field (e.g., \citealt{Page2007,Ruiz2010,Fauvet2011,Jaff2013,PlanckXLII2016,Pelgrims2018}). However, several studies have demonstrated that the magnetic field in the local ISM does not follow the large-scale magnetic field (\citealt{leroy_1999,Santos_2011,Frisch2012,berd2014}).

\cite{alves_2018} made the association between the local distortion of the magnetic field and the Local Bubble (LB), a local interstellar structure in the solar vicinity that likely results from supernova explosions (\citealt{cox_1982,cox_1987,Shelton1998,smith,Maiz-Apellaniz2001,Berghofer2002,fuchs,lallement_2003,lallement_2014,puspitarini,Liu2017,Sch2017}), and proposed a model of the magnetic field in the shell of the LB. They fit their model to the emission observed with \textit{Planck} that is polarized by dust at the Galactic polar caps where the contribution from the magnetized LB is \textit{\textup{expected}} to dominate the large-scale Galactic field. However, the possibility that the LB has no lids towards high-Galactic latitudes exists because it appears to be connected to surrounding cavities (\citealt{bailey_2016, lallement_2003,lallement_2014,Lallement2019,Leike2019,farhang_2019}).

It is therefore not clear whether the diffuse polarized emission, which is relevant for CMB studies and is observed at the Galactic caps, comes from local or more distant regions of the Galaxy.
This casts doubts on the attempts that have been made to constrain the magnetic field models, either at local or at large scales, which rely on these data sets.

Starlight polarization at the optical wavelength can help address this question. The polarized emission of dust at 353-GHz is due to aspherical interstellar dust grains that are aligned with the ambient Galactic magnetic field (e.g., \citealt{andersson_review} and references therein). The same dust grain population induces a net polarization to the incident initially unpolarized light that passes through a dusty region. Thus, the polarization position angles and polarization intensities are strongly correlated at the two observed frequencies \citep{martin_2007}. Based on this property, it is in principle possible to determine the line-of-sight distance of a dust-emitting region using starlight polarization as a distance candle.

In this letter we study the correlation properties between the polarization at the two frequencies as a function of distance in order to infer the level at which the LB contributes to the emission that is polarized by Galactic dust. Our aim is to determine (\textit{i}) if the high Galactic sky of thermal dust polarized emission is dominated by a nearby magnetized structure, and consequently, (\textit{ii}) if this emission can be used to constrain the local magnetic field.

\section{Archival data}

\subsection{Starlight-polarization data}
\label{subsec:starlight_data}

   \begin{figure*}[!h]
   \centering
    \includegraphics[width=\textwidth]{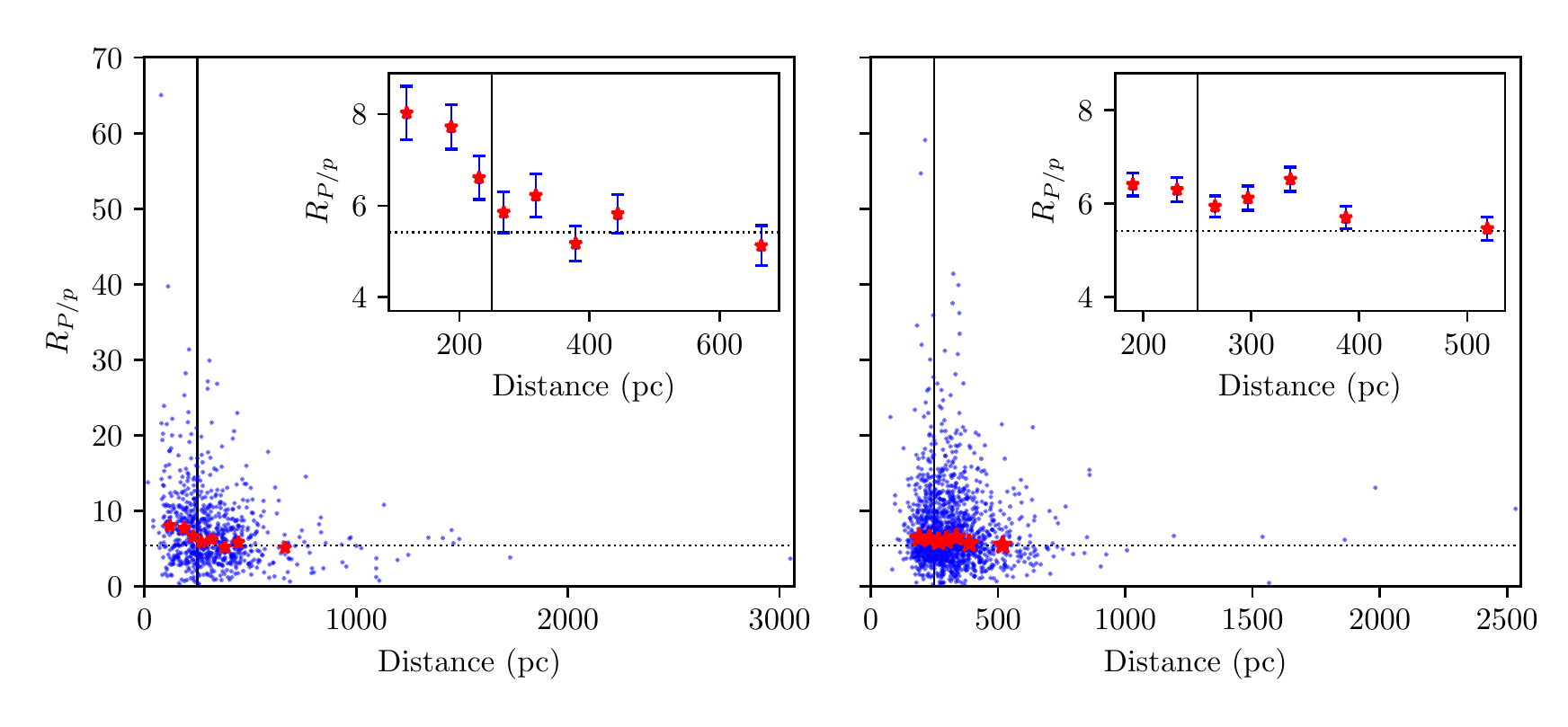}
      \caption{\textbf{Left:} $R_{P/p}$ vs. distance including lines of sight at high Galactic latitudes ($|b|>60^\circ$). The vertical solid line denotes $250$ pc. Each bin contains $100$ measurements, except for the last bin, which contains $89$. Red stars correspond to the median value of each distance bin. \textbf{Right:} Same as in the left panel, but for the intermediate-latitude lines of sight ($|b|<60^\circ$). Each bin contains $220$ measurements, except for the last bin, which contains $202$.}
         \label{fig:RPp_dist}
   \end{figure*}
   
We obtained the starlight-polarization data from the catalogs of \cite{berd2001_a}, \cite{berd2001_b}, \cite{berd2002}, and \cite{berd2014}. These catalogs are the largest polarization sample of the diffuse ISM at high Galactic latitudes, which are of interest for this study. All stars are located at latitudes $|b|>30^\circ$ , and the observations have been performed in the V band. To refer to optical polarization quantities, we thus use the subscript $\rm{v}$. There are $2881$ stars with measured degree of polarization, $p_{\rm{v}}$, uncertainty $\sigma_{p_{\rm{v}}}$ and polarization angle, $\psi_{\rm{v}}$, with uncertainty $\sigma_{\psi_{\rm{v}}}$, given in the IAU convention and in the celestial reference frame. We converted $\psi_{\rm{v}}$ into the Galactic coordinate system using Eq.~8 of \cite{panopoulou}, for instance, and finally determined the $q_{\rm{v}}$, $u_{\rm{v}}$ Stokes parameters in the HEALPix convention\footnote{\url{https://healpix.jpl.nasa.gov/html/intronode12.htm}} by 

    \begin{equation}
       \label{eq:stokes_parameters}
       q_{\rm{v}} =   p_{\rm{v}} \, \cos(2\psi_{\rm{v}})\,; \;
       u_{\rm{v}} = - p_{\rm{v}} \, \sin(2\psi_{\rm{v}})
    .\end{equation}
In this analysis we only considered starlight measurements with ${p_{\rm{v}}}/{\sigma_{p_{\rm{v}}}} \geq 2$, which corresponds to $2324$ stars. Whenever possible, we retrieved the star distances from \cite{bailer_jones}. This catalog includes distances implemented in a Bayesian approach from parallaxes of the \textit{Gaia DR2} release
\citep{gaia}. We adopted distances from the polarization catalogs for the stars that we could not cross-match in \cite{bailer_jones}. The parallaxes of these stars have been measured by Hipparcos. They are 519 in total, which is $22 \%$ of the total sample. Most of these stars are located at Galactic latitudes $|b| \geq 60 \degr$ and at distances between $100 - 300$ pc. We note, however, that the conclusions of this work remain consistent regardless of whether we use these measurements in our analysis.

\subsection{Submillimeter-polarization data}
\label{subsec:dust_emission_data}

We extracted submillimeter polarization for the lines of sight for which starlight data exist. To refer to submillimeter-polarization quantities, we use the subscript $S$.
We followed the process described in \cite{planck_2018} (see their Sect.~2: alternative Stokes maps production). We used the \textit{Planck} $353$-GHz $Q_{S}$, $U_{S}$ maps, multiplied by $287.5$ $\text{MJy sr}^{-1}\text{K}^{-1}_{\text{CMB}}$ in order to convert into astrophysical units, and finally smoothed the maps in order to increase the signal-to-noise ratio (S/N). We considered smoothing values of $10\arcmin$, $20\arcmin$, $30\arcmin$, $40 \arcmin$ , and $60\arcmin$ , which also allowed us to address the question of beam depolarization and beam difference (see Appendix \ref{app:varying_radii}). For our analysis we find that a smoothing radius of $20\arcmin$ is a good compromise between noisy data and large beam difference. We note, however, that the main conclusions reported in this work are found to be robust against this choice.
No correction was performed for the cosmic infrared background anisotropies or for the CMB polarized signal, which are both subdominant at $353$ GHz \citep{PlanckIV2018}.

For the per-pixel block-diagonal covariance matrix maps $C_{QQ}$, $C_{QU}$, and $C_{UU}$, we first converted into astrophysical units and then smoothed them according to the prescription in the appendix of \cite{planck_2015_appendix}. Then, we created the dust-polarized intensity map and its uncertainty using the equations

   \begin{equation}
      \label{eq:polarized_intensity}
      P_{S}          = \sqrt{Q_{S}^{2} + U_{S}^{2}}\, ; \; \sigma_{P_{S}}^{2} = \frac{ Q_{S}^{2}C_{QQ} + Q_{S}^{2}C_{UU} + 2Q_{S}U_{S}C_{QU}} {P_{S}}
   .\end{equation}
   
Both $p_{\rm{v}}$ and $P_{S}$ are biased quantities (e.g., \citealt{serkowski,montier1, montier2}). We used the modified asymptotic estimator of \cite{plaszczynski} to debias them.
In the case of the smoothing radius of $40\arcmin$ , our data sample perfectly agrees with the data set used in \cite{planck_2018}\footnote{V. Guillet, private communication} , which makes us confident in our post-processing of the \textit{Planck} data.

\section{Analysis}
\label{sec:RPp_distance}

We divided our sample into the "intermediate" ($30^\circ \leq |b| \leq 60^\circ$) and "high" ($|b| > 60^\circ$) Galactic latitudes and analyzed the two subsamples in parallel. The reason is that if the LB contributes to the submillimeter sky, then its imprint (if any) is more likely to be detectable at high Galactic latitudes because at lower latitudes, ISM matter from the Galactic disk dominates the line-of-sight integrated emission signal. The intermediate Galactic latitude subsample contains $1528$ lines of sight. The high-Galactic latitude subsample contains $796$ lines. 

\subsection{Emission-to-extinction ratio versus distance}
\label{subsec:median_RPp}

Our analysis relies on the assumption that the dust properties are the same throughout the Milky Way. For each line of sight we computed the emission-to-extinction polarization ratio as introduced in \cite{planck_2015} using the debiased quantities 

   \begin{equation}
     \label{eq:RPp}
     R_{P/p} = \frac{P_{S}}{p_{\rm{v}}}
   .\end{equation}
The units of $R_{P/p}$ are $\text{MJy sr}^{-1}$. The ratio quantifies how efficient the dust grains are in emitting polarized light compared to their ability of inducing optical polarization. When a star is located at infinity, $P_{S}$ and $p_{\rm{v}}$ trace the exact same dust material and $R_{P/p}$ approaches its mean value. The latter is a characteristic property of the population of aligned dust grains of our Galaxy. Its mean value, which we call "universal", was determined to be $R_{P/p} = 5.42 \pm 0.05 \text{MJy sr}^{-1}$ \citep{planck_2015, planck_2018}. In reality, $R_{P/p}$ varies with distance. For a given line of sight, $P_{S}$ accounts for the dust emission integrated up to infinity, whereas $p_{\rm{v}}$ accounts for dust extinction up to the distance of the star.

\smallskip

In Fig.~\ref{fig:RPp_dist} we show $R_{P/p}$ versus distance for the high (left panel) and intermediate (right panel) Galactic latitudes. In both cases we note that there are values up to $60\, \text{MJy sr}^{-1}$. As expected, this mostly occurs for nearby stars, while for the more distant stars, $R_{P/p}$ is closer to the universal value (dashed horizontal line). 
This trend is more prominent when we inspect the median values for individual distance bins, as is better shown in the insets of Fig.~\ref{fig:RPp_dist}. We binned both samples with distance by keeping the same number of measurements per bin. The median values are shown as red stars in Fig.~\ref{fig:RPp_dist}, and they are located at the median distance of stars at each bin. We propagated the observational uncertainties to the medians by creating $10^{4}$ mock data sets (see Appendix~\ref{app:uncertainties}) and taking the standard deviation of the distribution of the mock median $R_{P/p}$ in each distance bin. We observe a prominent difference between the high and intermediate Galactic latitudes. In the first case, $R_{P/p}$ approaches the universal value at $\sim 250$ pc and then remains constant\footnote{We emphasize the convergence of $R_{P/p}$ toward the universal value at nearby distances. Variations of this distance between different lines of sight are expected, but the current sample does not allow for such an investigation. In Appendix~\ref{app:sample_distribution} we show a potential signature of a chimney through the LB wall, however.}. On the other hand, in the intermediate-latitude case, the medians present a more complex behavior and remain at higher values. The values approach the universal $R_{P/p}$ value at larger distance ($\gtrsim 350$ pc).

In the high-latitude case, the monotonic convergence to the mean $R_{P/p}$ value indicates that independently of the line of sight, most of the total submillimeter polarization signal is captured within the first $250$ pc. This indicates the existence of an ISM dust wall at a distance of about $250$ pc with not much material beyond it. Otherwise, the medians could have taken any value at any distance.

\smallskip

\begin{figure}
    \includegraphics[trim={0cm 16.4cm 27.2cm 0cm},clip,width=\hsize]{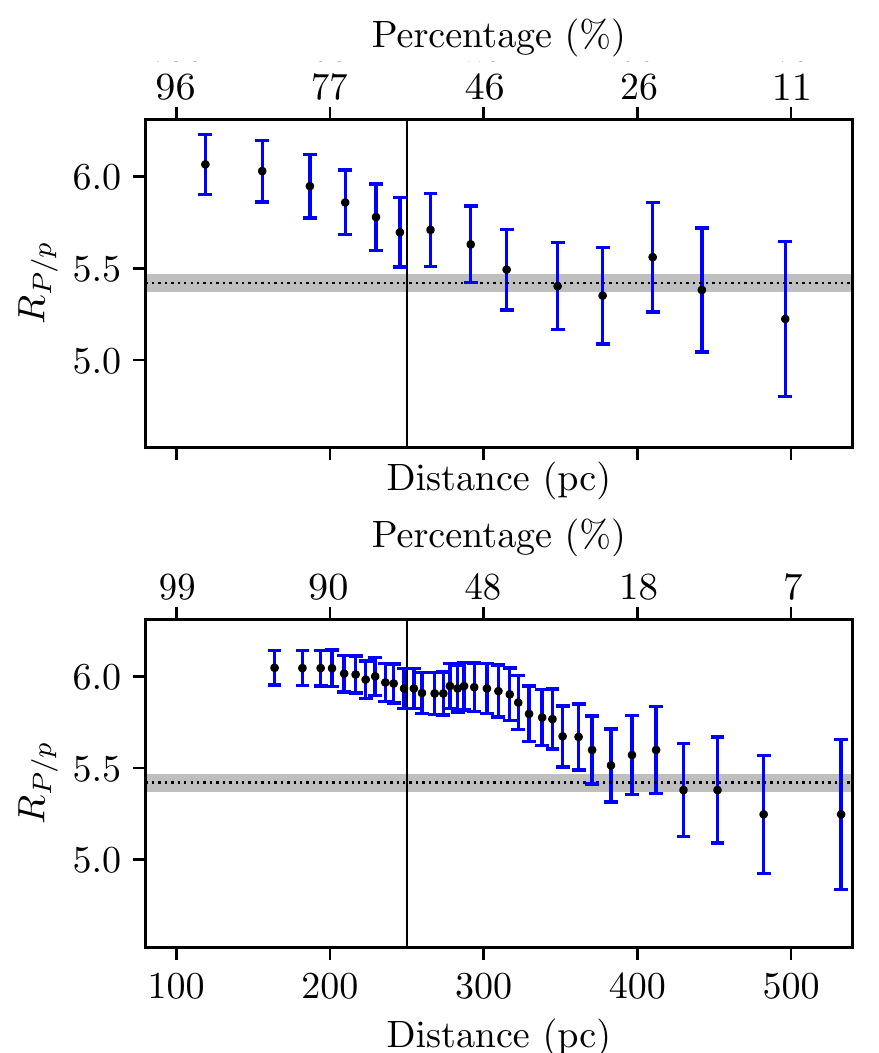}
      \caption{\textbf{Top panel:} Median $R_{P/p}$ of the high-latitude lines of sight as a function of the fraction of the truncated samples. Distances correspond to the minimum distance of each truncated sample in which we computed $R_{P/p}$. The dotted horizontal line corresponds to the mean $R_{P/p}$ value, $5.42 \text{MJy sr}^{-1} $, and the gray band to its one-sigma errors. \textbf{Bottom panel:} Same as above for the intermediate-latitude lines of sight.}
         \label{fig:RPp_truncated}
   \end{figure}

To infer this statement further in a bin-free way, we relied on a truncation analysis similar to \cite{pelgrims2019}.
We started with the full sample and computed the median of the $R_{P/p}$. We sorted the sample by star distances, gradually removed the nearest measurements, and computed the median $R_{P/p}$ for the truncated sample at each truncation step. In Fig.~\ref{fig:RPp_truncated} we show the $R_{P/p}$ medians as a function of the remaining sample size fraction (upper horizontal axis). Distances in the lower horizontal axis correspond to the minimum distance of each truncated sample. Inspecting the high-latitude lines of sight (top panel), we see again that the median $R_{P/p}$ approaches the universal value $5.42\, \text{MJy sr}^{-1} $ at $\sim 250$ pc. Here, it is evident that the nearby stars induce the high $R_{P/p}$ values because their optical polarization does not reflect the total dust column across a given line of sight. This indicates that there is significant dust material behind these stars. When these lines of sight are removed, the rest of the sample is consistent within $1\sigma$ with the universal value. For comparison we show the case of intermediate Galactic latitudes (bottom panel), which has a more complex behavior. The values are in general much higher than the value of the other case, and $R_{P/p}$ approaches the universal value at larger distance, $\gtrsim 350$ pc. The complex behavior is representative of a non-uniform ISM source and indicates the contribution of multiple polarizing layers in the observed signal.
Interestingly, the fact that we locate most of the dust material below $\sim 250$ pc and $\sim 350$ pc at high and intermediate Galactic latitudes, respectively, is consistent with a geometrical model of plane-parallel extra-planar dust with a thickness of about 240 pc \citep{van_Loon_2009}. The incompleteness of the current sample, however, does not allow us to further explore potential geometries and variations of the magnetized dusty structure (see Appendix~\ref{appfig:sky_distribution} for more details).

The comparison of Figs.~\ref{fig:RPp_dist} and~\ref{fig:RPp_truncated} allows us to conclude that most of the dust-polarization signal originates from a dusty magnetized ISM structure that extends roughly between 200 and 300 pc from the Sun, all lines of sight together. This statement is further inferred in Appendix~\ref{subsec:RPp_distance_statistics}, where we also test its statistical robustness.

\subsection{Correlation of the submillimeter and optical polarization angles versus distance}
\label{subsec:Psi_distance}

To explore further the likely dominant contribution of the LB to the observed dust-emission signal at large Galactic latitudes, we used the difference of polarization angles between optical and submillimeter frequencies.
We used the definition of \cite{planck_2015}
   \begin{equation}
       \label{eq:angles}
       \psi_{S/V} = 0.5 \times \arctan \left[(U_{S}q_{\rm{v}}-Q_{S}u_{\rm{v}}), -(Q_{S}q_{\rm{v}}+U_{S}u_{\rm{v}}) \right]
   ,\end{equation}
where $\psi_{S/V} \in [-90^\circ, 90^\circ]$. When optical and submillimeter polarization probe the same ISM dust across a given line of sight, the two polarization angles are perpendicular. Perfect orthogonality corresponds to zero $\psi_{S/V}$, while deviations about zero indicate that dust-emission and dust-extinction polarization trace different material. We expect that nearby stars that do not trace the total dust column density would have a $\psi_{S/V}$ distribution away from zero, whereas at larger distances, the distribution would approach zero. However, even for ideal cases, \cite{planck_2018} found a $-3.2^\circ$ systematic offset in the median of the $\psi_{S/V}$ distribution (see their Fig.~G.3). They also showed that this offset could be attributed to ISM background contribution that was not originally accounted for. More stringent criteria resulted in a median of $-2^\circ$. We consider this value as representative for the perfect correlation between submillimeter and optical polarization angles.

\begin{figure}
    \includegraphics[width=\hsize]{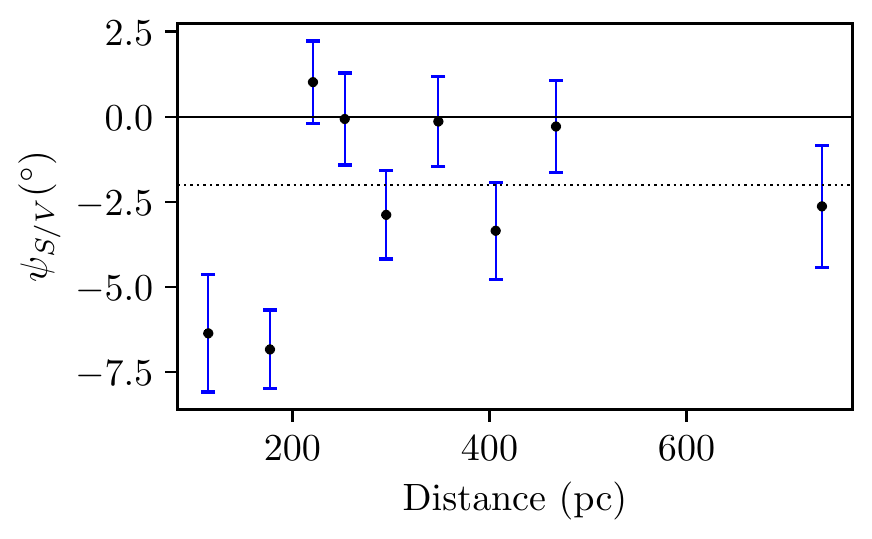}
      \caption{Difference in polarization angles, $\psi_{S/V}$, vs. distance at high Galactic latitudes. Each bin contains $90$ measurements, except for the last bin, which has $68$. We computed the errors on the medians using mock data sets from Appendix \ref{app:uncertainties}. The horizontal line corresponds to the $-2^\circ$ offset.}
         \label{fig:psi_sv}
   \end{figure}

In Fig.~\ref{fig:psi_sv} we visualize the binned $\psi_{S/V}$ versus distance for the high-Galactic latitude lines of sight as in Fig.~\ref{fig:RPp_dist}. The horizontal line is the $-2^\circ$ offset. It is evident that the low values of the nearby bins indicate that optical and submillimeter polarization do not trace the same material because at these distances, significant amount of dust lies behind the stars that optical polarization does not probe. At distances larger than $\sim 250$ pc, all the median values oscillate about $-2^\circ$ , indicating that we reach the perfect case where both observables trace the same ISM dust. 

\section{Discussion and conclusions}

Several studies have explored the magnetic field in the solar neighborhood through samples of optical polarization through starlight (\citealt{leroy_1999, Andersson2006,Santos_2011,frisch,Medan2019,Cotton2019}). All agree that the local magnetic field does not follow the large-scale Galactic field, and all make connections with the LB. \cite{Gontcharov2019} showed that the polarized intensity of stars reaches a plateau at a distance between $150$ -- $250$ pc (all azimuths together), a trend that is also observed in the ratio of extinction up to the star distance over the total extinction. \cite{clark_2014} compared starlight-polarization angles with the orientation of H$_{\rm{I}}$ fibers and inferred their line-of-sight distances. They reported that the majority of these structures lie in a similar range of distances within the LB. These conclusions are in line and support our finding.

The scope of our work was to investigate the contribution of the LB to the integrated submillimeter polarized emission. Using optical polarization data as distance candles, we inferred the dust-emission properties at high Galactic latitudes. We studied the behavior of the $R_{P/p}$ ratio versus distance for lines of sight at high and intermediate Galactic latitudes. We find evidence that at high latitudes, a polarizing layer exists. This structure contributes significantly to the total observed submillimeter-polarization signal. With the current sample it is not possible to locate the exact edges, and such an investigation would be beyond the scope of the current work.
On the other hand, the intermediate-latitude lines of sight are more consistent with the existence of multiple polarizing sources located at various distances. We extracted similar evidence by comparing the polarization angles from the optical and the submillimeter, $\psi_{S/V}$, as a function of distance.
Our sample is biased toward lines of sight in the northern hemisphere. However, we expect that our conclusions hold for the southern hemisphere as well (see discussion in Appendix~\ref{app:sample_distribution}).

The distance at which our investigations place the main region of dust-polarization emission is consistent with the thick shell of the LB toward the Galactic caps as revealed by X-ray observations \citep{puspitarini,Liu2017} and recent three-dimensional maps of the local ISM \citep[e.g.,][]{Lallement2019,Leike2019} and the aforementioned estimates. This makes us confident in our claim that the high-latitude dust-polarized emission as observed by \textit{Planck} at $353$-GHz is dominated by the magnetized LB. We therefore conclude that toward the polar caps, dust-polarized emission can be used to constrain the modeling of the local magnetic field. As a corollary, we find that the modeling of the local magnetic field, and therefore of the LB, is of paramount importance for the characterization of the high-frequency Galactic foregrounds of CMB polarization.

We used a limited number of lines of sight mostly toward the northern hemisphere, and most of the stars we used are located nearby. These are caveats that we cannot overcome due to the limited amount of the existing data. Upcoming polarization surveys, such as PASIPHAE \citep{tassis_pasiphae}, will provide optical polarization measurements for millions of stars at Galactic latitudes $|b|>60^\circ$ in both hemispheres. This will enable a direct and bias-free comparison between optical- and submillimeter-polarization data and allows for a tomographic decomposition of the dust-polarized emission. Such a large-scale survey will be valuable in the context of CMB experiments and large-scale magnetic field studies, and we look forward to carry out the present analysis with such an upgraded data set.

\begin{acknowledgements}
We thank K. Tassis for his continued support during the elaboration of this work. We thank V. Guillet for providing us with the data sample used in \cite{planck_2018} with which we started this project before we compile our sample and for helpful comments. We also thank G. V. Panopoulou for useful comments on our analysis and J. van Loon for his fruitful review. RS would like to thank Dr K. Christidis for fruitful discussions. This project has received funding from the European Research Council (ERC) under the European Union’s Horizon 2020 research and innovation program under grant agreement No 771282. We acknowledge the use of data from the Planck/ESA mission, downloaded from the Planck Legacy Archive, and of the Legacy Archive for Microwave Background Data Analysis (LAMBDA). Support for LAMBDA is provided by the NASA Office of Space Science. Some of the results in this paper have been derived using the HEALPix package \citep{Gorski2005}. This work has made use of data from the European Space Agency (ESA) mission Gaia (https://www.cosmos.esa.int/gaia), processed by the Gaia Data Processing and Analysis Consortium (DPAC, https://www.cosmos.esa.int/web/gaia/dpac/consortium). Funding for the DPAC has been provided by national institutions, in particular the institutions participating in the Gaia Multilateral Agreement. 
\end{acknowledgements}

\bibliographystyle{aa}
\bibliography{bibliography}

\begin{appendix}

\section{Beam depolarization and smoothing radius}
\label{app:varying_radii}
There is a beam difference between optical and submillimeter data sets involved in this work (see Fig.~1 of \cite{planck_2015} for a sketch illustrating the difference in geometry). The resolution of the 353 GHz \textit{Planck} observation is limited by the instrument beam ($4.94\arcmin$), which is larger than that of optical polarization observations, for which pencil beam applies. As a result, different polarization sources contribute within a \textit{Planck} beam and average out their signal. This is a beam-depolarization effect that induces a negative bias to the observed signal (for more details, see Appendix~G.1c of \cite{planck_2018}). Hence there is a beam difference between the two frequency channels that are used.
For a given line of sight, submillimeter polarization traces a greater volume of ISM material than is traced by the optical polarization. This difference becomes more prominent at larger distances.
This effect is due to native resolution, and an additional smoothing of the diffuse emission is necessary for polarization channels in order to increase the signal-to-noise ratio (S/N) of the data.
This smoothing further averages out the signal from different polarization sources and increases the effective beam of the data. In order to compare the two observables, we would ideally use the smallest possible radius for the smoothing. However, in this case, the S/N in the dust emission data would be very low. The compromise is to mitigate the beam difference and beam depolarization and work with sufficiently high quality data.

\begin{figure}
   \centering
   \includegraphics[width=\hsize]{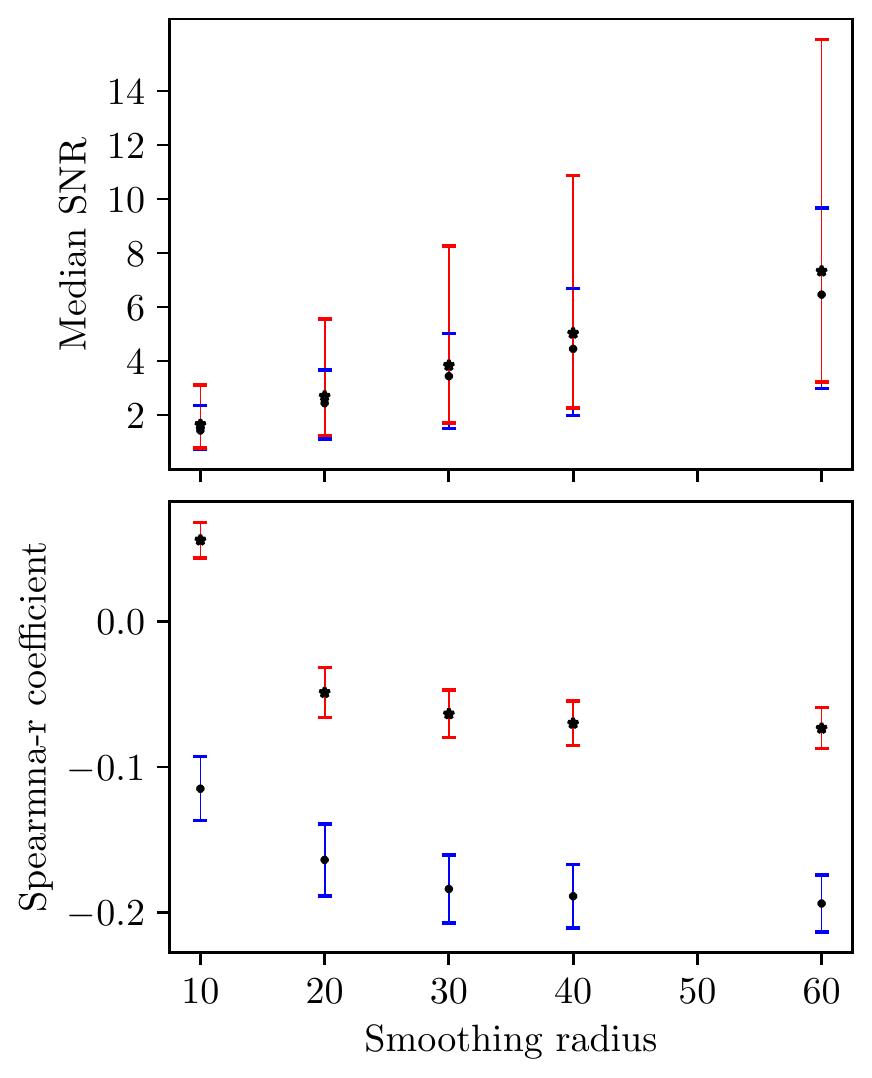}
      \caption{Asterisk markers with red error-bar lines correspond to the intermediate-latitude sample, and dot markers with blue error-bar lines to the high-latitude sample.
      \textbf{Top panel:} Median S/N of $P_{S}$ as a function of smoothing radius. The error bars correspond to the $68\%$ percentile of the distribution.
      \textbf{Bottom panel:} Mean Spearman r-coefficient derived from a distribution using mock data as a function of the smoothing radius. The 1$\sigma$ error bars correspond to propagated observational uncertainties.}
     \label{fig:varying_smoothing_radius}
   \end{figure}

To infer the effect of the above systematic on our results and choose the optimal smoothing radius, we tested our analysis using different radii to smooth the \textit{Planck} maps (Sect.~\ref{subsec:dust_emission_data}). In the top panel of Fig.~\ref{fig:varying_smoothing_radius}, we show the median S/N of $P_{S}$ as a function of the smoothing radius. The error bars  correspond to the $68\%$ percentile. A linear correlation between the S/N of $P_{S}$ and the smoothing radius of the \textit{Planck} maps is prominent.
In the bottom panel we present the mean Spearman r-coefficient as a function of the smoothing radius. For each radius we created a Spearman r-coefficient distribution of the $R_{P/p}$ distance relation (see Sect.~\ref{subsec:RPp_distance_statistics}) using $10^{4}$ mock data-sets as described in Appendix~\ref{app:uncertainties} to propagate the observational uncertainties of $R_{P/p}$ down to the measure of the correlation coefficient. 
The error bars shown are the 1$\sigma$ deviation about the mean.
For $20\arcmin$ and above, the mean values do not change significantly, even though the S/N increases. This indicates that the inherent correlation is captured even at $ 20\arcmin$. Complementary, the median S/N of $P_{S}$ at this radius is $\sim 2.4$, which satisfies the threshold we used in optical polarization data (Sect~\ref{subsec:starlight_data}). This is a good compromise between high-quality data and using a small smoothing radius. If we were to use $10\arcmin$, the median S/N could be 1.4. Consequently, the correlation between $R_{P/p}$ and distance would be smeared out by the noise and would not be significant. In addition, when we produce uncorrelated realizations of the data, shown by the green distribution in Fig~\ref{fig:RPp_shuffled}, and compare it with the observed Spearman r-coefficient distribution, shown by the brown distribution in the same figure, we find a significant overlap for the case of the $10\arcmin$ radius. This means that the probability of reproducing the observed correlation from a random sample is high. This probability is significantly reduced for radii $20\arcmin$ and larger. Altogether, the fact that the Spearman r-coefficient remains constant for radii $\geq 20$ and the significant increase of the median S/N of $P_{S}$ give us confidence to adopt $20\arcmin$ as the best-fit value to process the \textit{Planck} data.

\section{Propagation of observational uncertainties}
\label{app:uncertainties}
We rely on the generation of mock observations through Monte Carlo simulations to propagate the observational uncertainties on the polarization quantities in our analysis.
For the optical polarization by starlight, we assume that the Stokes parameters have Gaussian errors and that $\sigma_{q_{\rm{v}}}\simeq\sigma_{u_{\rm{v}}}\simeq\sigma_{p_{\rm{v}}}$, which is valid for typical $p_{\rm{v}}$ values in the ISM \citep{serkowski}.
Let $N$ be observational pairs of Stokes $(q_{v,i}, u_{v,i})$, where $i=1, \dots, N$.
From each $q_{v,i}$ and $u_{v,i}$ , we draw $10^{4}$ values according to a Gaussian distribution with standard deviation $\sigma_{p_{v,i}}$. We obtain $10^{4}$ mock data-sets that each consist of $N$ $(q_{mock,i}$, $u_{mock,i})$ pairs. 

For the thermal dust polarization at submillimeter wavelength, we consider the noise correlation in the Stokes parameters as they are given \citep{planck_2018}. Let $(Q_{S,i}, U_{S,i})$ be the observed Stokes parameters, where $i=1, \dots, N$ is the index of the observational pair. The noise correlation of the Stokes parameters is described by the elements $C_{QU,i}, C_{IQ,i}, \text{and } C_{IU,i}$. We followed Appendix A.1.3 of \cite{planck_2015_appendix} to generate correlated noise, $[N]_{i}$, that reflects the covariance matrix through Cholesky decomposition $[C]_i = [L]_i^\dagger \times [L]_i$ , where $[L]_{i}$ is a low-triangular matrix. We have
\begin{equation}
    [N]_{i} = [L]_{i} \times [G]_{i}
,\end{equation}
where $[G]_{i}$ is a random realization of the normal distribution. $[N]_{i}$ is then added to each of the observed Stokes parameters. The mock data are consequently extracted using 
\begin{gather}
\begin{pmatrix}
I_{mock,i} \\
Q_{mock,i} \\
U_{mock,i} \\
\end{pmatrix}
=
\begin{pmatrix}
I_{obs,i} \\
Q_{obs,i} \\
U_{obs,i} \\
\end{pmatrix}
+
[L]_{i} \times [G]_{i.}
\end{gather}
We assume that $I$ is perfectly known, implying $C_{II}=C_{IQ}=C_{IU}=0$. This is a reasonable assumption because the S/N in $I$ is orders of magnitude higher than in $Q$ and $U$ for \textit{Planck} $353$ GHz data. Then, the Cholesky matrix reduces to
\begin{gather}
[L]_{i}
=
\begin{pmatrix} 
0  &  0                             &  0  \\
0  &  \sqrt{C_{QQ,i}}                 &  0  \\
0  &  \frac{C_{QU,i}}{\sqrt{C_{QQ,i}}}  &  \sqrt{C_{UU,i} - \frac{C^{2}_{QU,i}}{C_{QQ,i}} }
\end{pmatrix}
\end{gather}
We thus obtain the mock Stokes parameters by
\begin{align}
    & Q_{mock,i} = Q_{obs,i} + \sqrt{C_{QQ,i}} \times G_{Q,i} \\ 
    & U_{mock,i} = U_{obs,i} + \Bigg( \sqrt{C_{UU,i}-  \frac{C_{QU,i}^{2}}{C_{QQ,i}}} + 
                   \frac{C_{QU,i}}{\sqrt{C_{QQ,i}} } \Bigg) \times G_{U,i,}
                   \; .
\end{align}
where $G_{Q,i}$ and $G_{U,i}$ are random numbers drawn from normalized Gaussians. We draw $10^{4}$ numbers from each $G_{Q,i}$, $G_{U,i}$ and obtain $10^{4}$ different mock data-sets that each consist of $N$ $(Q_{mock,i}$, $U_{mock,i})$ pairs.

We then combine the $10^{4}$ mock samples of starlight polarization and of dust polarization and reproduce the analyses as with the observed sample. We determine the observational uncertainties on our measurements by taking the standard deviation through the $10^{4}$ measurement realizations.

\section{Statistical tests on the $R_{P/p}$-distance relation}
\label{subsec:RPp_distance_statistics}
In this appendix we aim to answer the question whether for the high-latitude lines of sight \textit{\textup{1) the correlation between $R_{P/p}$ and distance is significant. We also determine  2) if it might be attributed to a random statistical process.}}

\begin{figure}
    \includegraphics[width=\hsize]{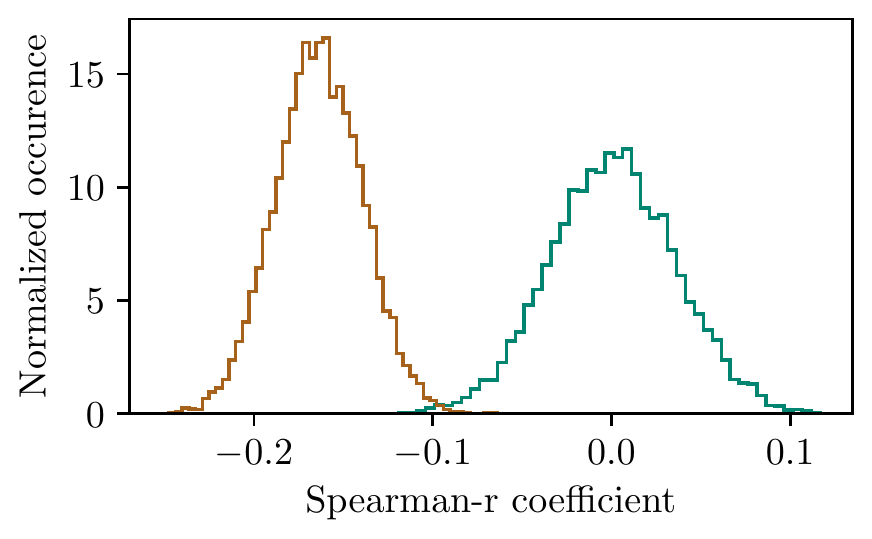}
      \caption{In brown we show the Spearman r-coefficient distribution of the mock data-sets. The distribution of the shuffled data is shown in green.}
         \label{fig:RPp_shuffled}
   \end{figure}

In order to answer the first question, we made the null hypothesis that there is no correlation between $R_{P/p}$ and distance, and we performed a Spearman r-test by setting the significance level at $0.01 \%$. The Spearman rank coefficient of our measurements is $-0.21,$ with a p-value of $10^{-7} \%$, which indicates that we can reject the null hypothesis with great confidence. In addition, we tested the contribution of the observational uncertainties in the Spearman test. We created $10^{4}$ mock data-sets (see Appendix~\ref{app:uncertainties}). For each data-set we computed the Spearman r-value, and we created the distribution that is shown in brown in Fig.~\ref{fig:RPp_shuffled}. When observational uncertainties are accounted for, the $R_{P/p}$ distance correlation is significantly different from zero.

In order to answer the second question, we performed a permutation test: we shuffled the distances on the $R_{P/p}$ values and computed the Spearman rank coefficients of the shuffled data. Using $10^4$ such shuffled data samples, we created the distribution corresponding to randomness. In Fig.~\ref{fig:RPp_shuffled} we show the distribution of the Spearman rank coefficients of the shuffled data in green. The distribution is centered at zero, which is characteristic of a random process with no correlation. The green and brown distributions overlap by $3\%$ . 

\begin{figure}
    \includegraphics[width=\hsize]{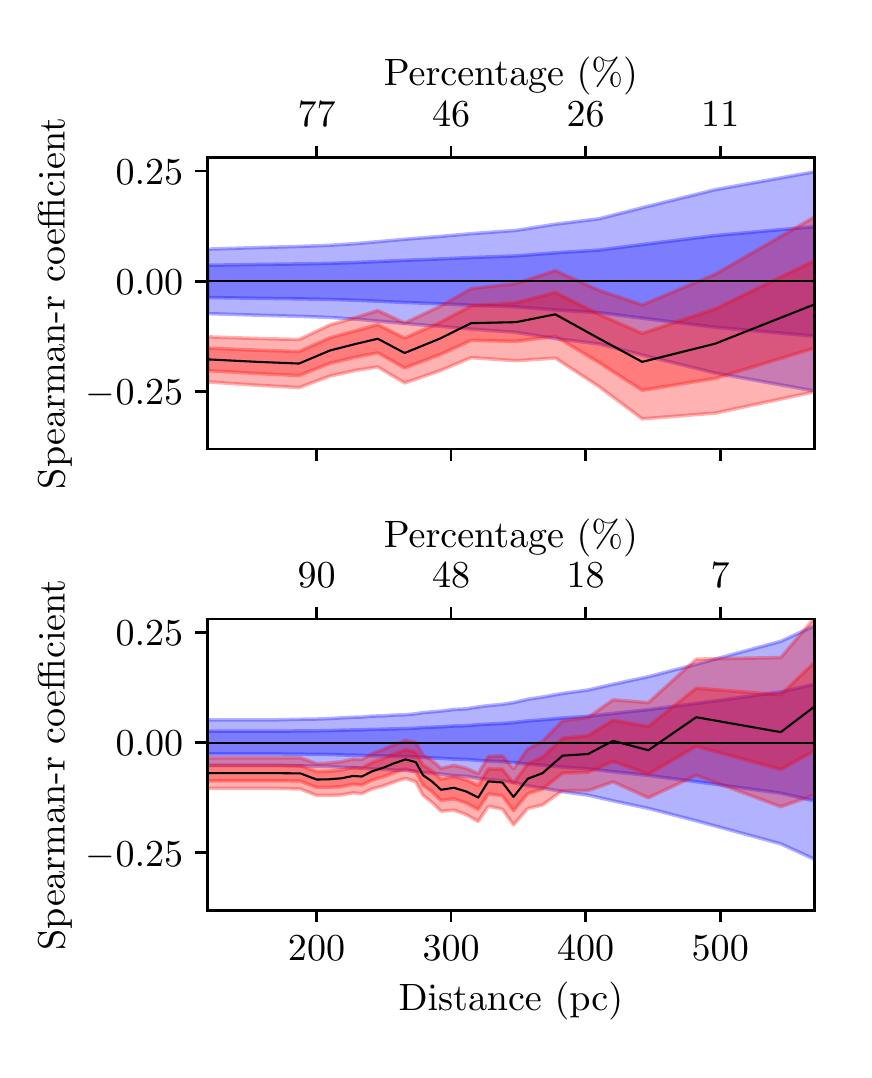}
      \caption{\textbf{Top panel:}  Spearman rank coefficient of the high-latitude lines of sight as a function of the truncated samples is shown with the solid  black curve. Distances 
               correspond to the minimum distance of each truncated sample in which we compute the correlation coefficient. The red shaded regions correspond to the 1 and 2$\sigma$ 
               deviations, respectively. The blue shaded regions correspond to 1 and 2$\sigma$ 
               deviations of the uncorrelated case, respectively. \textbf{Bottom panel:} Same for the 
               intermediate-latitude sample.}
         \label{fig:Spearman_truncated}
   \end{figure}

In Fig.~\ref{fig:Spearman_truncated} we visualize the reverse cumulative of the Spearman rank coefficient. We computed the coefficient of the total sample and then gradually removed measurements starting from the nearby ones, as in Sect.~\ref{subsec:median_RPp}. The black curved line shows the correlation coefficient as a function of the fraction of the truncated sample (upper horizontal axis) and distance (lower horizontal axis). We used mock data (Appendix \ref{app:uncertainties}) to evaluate the uncertainty of each truncated sample. The 1 and 2$\sigma$ deviations about the mean are shown as red shaded regions. We repeated this for the shuffled data. The Spearman coefficient of the shuffled data is constantly at zero (no correlation), and the blue shaded lines correspond to 1 and 2$\sigma$ about zero. The black solid line shows that when we truncate the sample, the correlation of the sample decreases. At $\sim 250$ and $\sim 300$ pc the correlation is consistent within $2$ and $1$ $\sigma$ with the uncorrelated case, respectively. This means that when the nearby measurements are removed, the observed convergence of $R_{P/p}$ becomes less prominent. Above a threshold located at about $300$ pc, $R_{P/p}$ and distance are no longer correlated. This is consistent with the picture we observe in Fig.~\ref{fig:RPp_dist} and~\ref{fig:RPp_truncated}. When the full submillimeter-polarization signal is captured, that is, when the sample is located between the stars and the observer, the sample becomes insensitive to the distance. The optical polarization by starlight and the submillimeter-polarized emission trace the same ISM material. For comparison we show the same plot for the case of lines of sight at intermediate Galactic latitudes. The correlation here, if any, is much shallower and could be attributed to a random statistical process at the 2$\sigma$ level. This case is more consistent with a picture in which different ISM polarizing sources contribute to the observed signal.

\section{Sample distribution on the sky}
\label{app:sample_distribution}

We explored variations of the polarization ratio $R_{P/p}$ as a function of sky coordinates in order to identify non-uniformities of the LB structure. In Fig.~\ref{appfig:sky_distribution}
we present an orthographic projection of $R_{P/p}$. For completeness, we show the $P_{S}$, $p_{\rm{v}}$ and star distances maps because variations in these maps propagate to $R_{P/p}$.

 Figure~\ref{appfig:sky_distribution} shows that in general, $R_{P/p}$ is more uniform toward longitudes, $l$, from $240 \degr$ to $60 \degr$. We focused on a special feature close to the northern Galactic cap, $b \geq 75 \degr$, and at longitudes $240 \degr \leq l \leq 300 \degr$. An $R_{P/p}$ excess is visible there. We compare this region with the other panels and note that the stars there are located at distances larger than $300$ pc. In this region, optical polarization has typical values of $p_{\rm{v}} \sim 0.15 \%$, while $P_{S}$ shows an excess, $\sim 3 \text{MJy sr}^{-1}$. This represents the case where submillimeter polarization traces more dust material across a given line of sight than in the optical. This is indicative of an LB chimney scenario toward this direction, which appears to be consistent with the 3D dust extinction maps (see \citealt{Lallement2019}, panel 9 in Fig.~12 and panel 1 in Fig.~13 for a qualitative comparison). Other features also deserve our attention, but we cannot conclude about them with the current biased sample.

Most of the measurements lie toward the northern hemisphere. This implies that our conclusion is driven by the northern Galactic hemisphere. However, the LB wall
shows prominent chimneys in the north when we consider the 3D extinction maps (e.g., \citealt{Lallement2019}). On the other hand, in the southern hemisphere the 3D extinction maps are more uniform and a continuous wall is more favorable, although there are chimney indications based on the 3D diffuse interstellar-band maps (\citealt{bailey_2016,farhang_2019}). Therefore we argue that our conclusions should hold in the south as well. Future data sets of optical polarization of stars will help verify this statement.

\begin{landscape}
\begin{center}
\begin{figure}
\centering
\includegraphics[trim={0cm 10.5cm 0cm 0cm},width=1.3\textwidth,clip,angle=0]{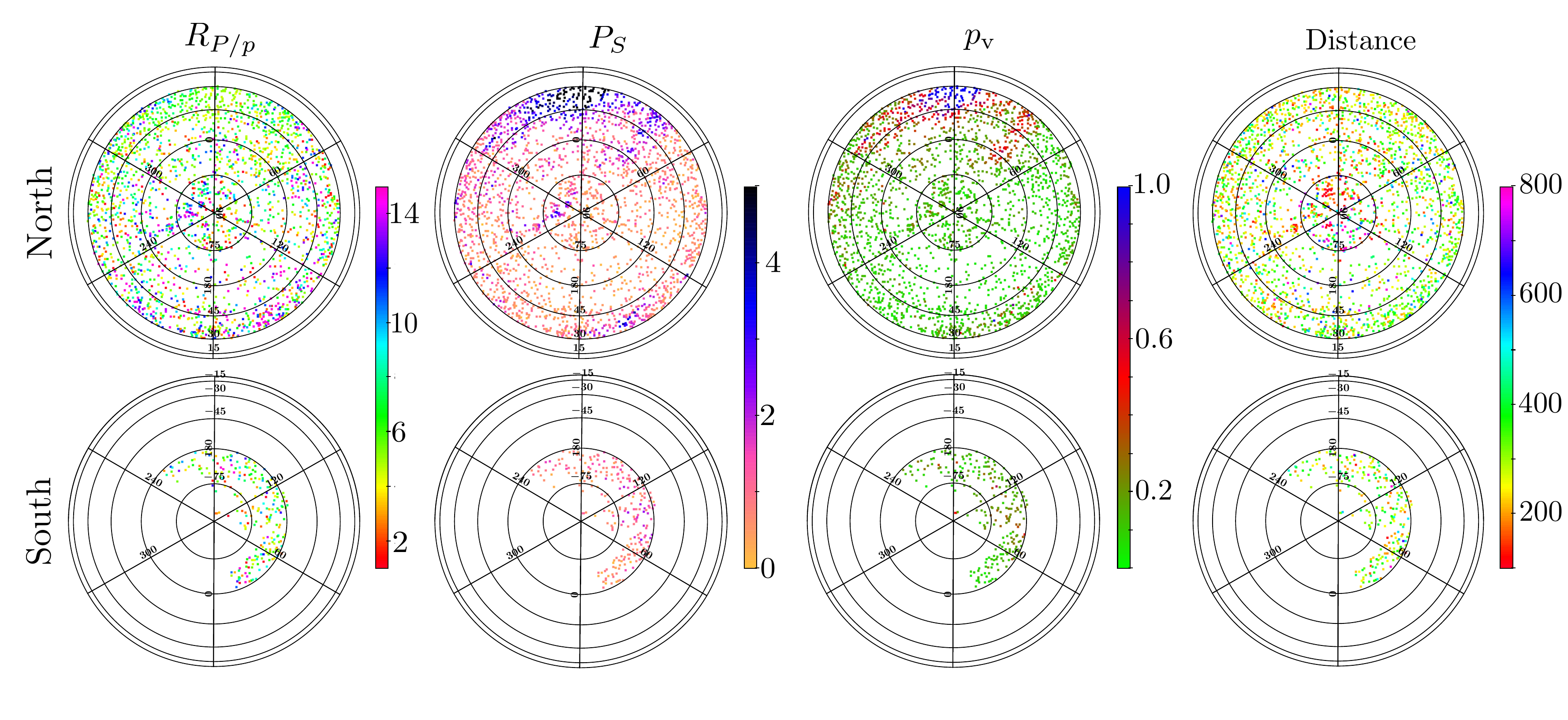}
\caption{Polar view of the sample of stars (marked as points) with color-coded quantities: $R_{P/p}$, $P_{S}$, $p_{\rm{v,}}$ and distance. The units of $R_{P/p}$ and $P_{S}$ are 
$\text{MJy sr}^{-1}$. The degree of polarization, $p_{\rm{v}}$, is measured in percentage ($\%$) and the distance in parsec. The northern hemisphere is shown in the top row and 
the southern in the bottom row. For $R_{P/p}$, only stars with $P_S/\sigma_{P_S} \geq 2$ are shown, and in the other maps all measurements are displayed. Several large-scale structures can be identified because they present high polarization, $p_{\rm{v}}$ and $P_{S}$ . Figure created with TOPCAT \citep{topcat}.}
\label{appfig:sky_distribution}
\end{figure}
\end{center}
\end{landscape}

\end{appendix}

\end{document}